\documentclass[12pt]{article}

\usepackage{graphicx,epstopdf}
\usepackage{amsmath}
\usepackage{epsf}

\def\be{\begin{equation}}
\def\ee{\end{equation}}
\def\ba{\begin{eqnarray}}
\def\ea{\end{eqnarray}}
\def\ge{\mathrel{\raise.3ex\hbox{$>$\kern-.75em\lower1ex\hbox{$\sim$}}}}
\def\la{\mathrel{\raise.3ex\hbox{$<$\kern-.75em\lower1ex\hbox{$\sim$}}}}

\def\simgt{\mathrel{\raise.3ex\hbox{$>$\kern-.75em\lower1ex\hbox{$\sim$}}}}
\def\simlt{\mathrel{\raise.3ex\hbox{$<$\kern-.75em\lower1ex\hbox{$\sim$}}}}

\newcommand{\bi}[1]{\bibitem{#1}}
\newcommand{\fr}[2]{\frac{#1}{#2}}

\newcommand{\nc}{\newcommand}

\nc{\gone}{\bar g_{\pi NN}^{(1)}}
\nc{\gzero}{\bar g_{\pi NN}^{(0)}}
\nc{\al}{\alpha}
\nc{\ga}{\gamma}
\nc{\de}{\delta}
\nc{\ep}{\epsilon}
\nc{\ze}{\zeta}
\nc{\et}{\eta}
\nc{\Th}{\Theta}
\nc{\ka}{\kappa}
\nc{\rh}{\rho}
\nc{\si}{\sigma}
\nc{\ta}{\tau}
\nc{\up}{\upsilon}
\nc{\ph}{\phi}
\nc{\ch}{\chi}
\nc{\ps}{\psi}
\nc{\om}{\omega}
\nc{\Ga}{\Gamma}
\nc{\De}{\Delta}
\nc{\La}{\Lambda}
\nc{\Si}{\Sigma}
\nc{\Up}{\Upsilon}
\nc{\Ph}{\Phi}
\nc{\Ps}{\Psi}
\nc{\Om}{\Omega}
\nc{\ptl}{\partial}
\nc{\del}{\nabla}
\nc{\ov}{\overline}
\nc{\newcaption}[1]{\centerline{\parbox{15cm}{\caption{#1}}}}

\begin{document}

\begin{titlepage}

\rightline{CERN-PH-TH/2006-098}
\rightline{hep-ph/0610003}

\setcounter{page}{1}

\vspace*{0.2in}

\begin{center}

\hspace*{-0.6cm}{\Large \bf Electric dipole moment constraints on \\ minimal electroweak baryogenesis}

\vspace*{0.5cm}
\normalsize

{\bf Stephan J. Huber$^{\,(a)}$, Maxim Pospelov$^{\,(b,c)}$ and
Adam Ritz$^{\,(b)}$}

\smallskip
\medskip

$^{\, (a)}${\it Theoretical Division, Department of Physics, CERN,
    Geneva 23, CH-1211 Switzerland}

$^{\,(b)}${\it Department of Physics and Astronomy, University of Victoria, \\
     Victoria, BC, V8P 5C2 Canada}

$^{\,(c)}${\it Perimeter Institute for Theoretical Physics, Waterloo,
Ontario N2J 2W9, Canada}

\smallskip
\end{center}
\vskip0.2in

\centerline{\large\bf Abstract}

We study the simplest generic extension of the Standard Model which allows for conventional
electroweak baryogenesis, through the addition of dimension six operators in the Higgs sector.
At least one such operator is required to be $CP$-odd, and we study the constraints on
such a minimal setup, and related scenarios with minimal flavor violation, from 
the null results of searches for electric dipole moments (EDMs), utilizing
the full set of two-loop contributions to the EDMs. The results indicate that the
current bounds are stringent, particularly that of the recently updated neutron EDM,
but fall short of ruling out these scenarios.  The next generation of EDM experiments should be sufficiently sensitive 
to provide a conclusive test.

\vfil
\leftline{September 2006}

\end{titlepage}

\section{Introduction}

The existence of a mechanism for electroweak baryogenesis (EWBG) \cite{Kuzmin} is one of the most remarkable and indeed elegant 
features of the Standard Model (SM), combining as it does various subtle aspects of the electroweak sector of the theory. Of course,
the fact that the Standard Model could in principle support electroweak baryogenesis now appears to be merely a mirage
 -- the $CP$-violation induced by the CKM phase is apparently many orders of magnitude
too small \cite{ckm}, and the Higgs is too heavy to allow for a sufficiently strong first-order electroweak phase transition \cite{KLRS96}.

A more optimistic viewpoint is that this failure of the Standard Model is a hint 
toward the presence of new physics. For example, the minimal supersymmetric standard model (MSSM) 
may have enough flexibility to ameliorate 
both deficiencies of the SM. 
Indeed, the MSSM still allows for electroweak baryogenesis, albeit in what is now a 
rather tuned region of the parameter space \cite{MSSM}. The null results of electric dipole moment (EDM) searches 
impose quite stringent
constraints on the spectrum if one requires  access to order-one $CP$-odd phases from the soft-breaking
sector. Combined with the requirement of an additional light scalar to afford a sufficiently strong first-order
transition, this leads to an ``almost split'' spectrum, with a single additional light scalar degree of freedom,
the right-handed stop. The EDMs are then those of split SUSY \cite{split}, and still allow for an order-one phase in the chargino
sector. Of course, this spectrum is seemingly rather tuned.

A currently popular alternative is the use of leptogenesis \cite{FY86,BPY05}, which divorces baryogenesis from 
the electroweak scale,
and utilizes new $CP$-odd phases in the lepton sector. Leptogenesis appears perfectly viable, and will receive
a considerable boost if indeed neutrinos are found to be Majorana, as this would strongly motivate new physics at
the see-saw scale. Unfortunately,  leptogenesis is currently, and may remain for some time, 
very difficult to test. EDMs are, at least without significant additional assumptions, relatively unaffected
by new Majorana phases in the lepton sector. Indeed, in the most minimal see-saw scenario, the additional
contribution is lost orders of magnitude below the already tiny Standard Model contribution to e.g. the 
electron EDM. This suppression is easily understood as EDMs do not violate lepton number, and so 
$\De d_e\propto G_F^2m_e m_{\nu}^2$ \cite{majoranaEDM}. 

Given this current state of affairs, and with the LHC hopefully going to illuminate the Higgs sector in the
near future, it seems appropriate to reconsider the status of electroweak baryogenesis in a more general context.
Several groups have recently taken a more general effective field theory approach to the new physics required
in order for EWBG to be viable \cite{Z93,ZLWY94,GSW04,HO04,BFHS04}. The minimal possibility is the addition of 
two dimension-six operators
to the Higgs sector, e.g.
\be
 {\cal L}_{\rm dim-6} = \frac{1}{\La^2} (H^\dagger H)^3 + \frac{Z_t}{\La^2} (H^{\dagger} H) t^c H Q_3,
  \label{L1}
\ee
where $\La$ denotes the scale of the new physics threshold. The first operator here serves to strengthen the
first-order transition, while the second allows for a new $CP$-odd phase in the coupling to the top-quark.
This simple modification was found to allow for a viable $\et_b\sim 10^{-10}$, provided the new
threshold was in the range $\La \sim 500-1000$ GeV.

There are several questions one may raise concerning such a new threshold. Firstly, the full set of allowed
dimension-six operators is very large, and indeed many require percent 
level tuning of Wilson coefficients 
if generated at such a low threshold, even in the absence of
new flavor structures; e.g. the oblique corrections from operators such as $|H^{\dagger}D_\mu H|^2$ would
generically be far too large. However, if $Z_t$ is promoted to be flavor-diagonal $Z^u_{ij} = Z^u Y^u_{ij}$,
then the tuning of these operators that is required is generally no worse than a few percent, and indeed
not significantly different from the tuning needed for EWBG in the MSSM. This could presumably be
further ameliorated with additional symmetries, restricting the generation of dangerous operators 
to loop level\footnote{A simple way to obtain the operators in (\ref{L1}) is to integrate out a weak scale
gauge-singlet scalar field. Oblique corrections and flavor violation are suppressed in this case \cite{GSW04}.}. 
Another dangerous class of flavor-diagonal operators, the EDM operators themselves,
need to be forbidden at up to two loops, which is the level at which they will be regenerated by the operator in (\ref{L1}),
as we will discuss in more detail.

The second, and perhaps more pertinent, question one may raise concerns the general viability of the scenario. 
Indeed, do the EDM bounds really allow new order-one $CP$-odd sources at such a low threshold? This is the question we would like to address in 
this work. Indeed, certain constraints were not considered in full in the preceding work, particularly color EDM contributions
to the hadronic EDMs, which do have a significant bearing on the conclusions. However, we find that although the constraints are
strong they currently fall short of excluding these scenarios. The current status is such that the next generation
of EDM experiments will however provide a conclusive test, as they will for EWBG in the MSSM.

The layout of this note is as follows. In the next section, we discuss the set of additional operators
we will consider, and also comment on the tuning inherent in ignoring other dimension-six terms at the
threshold. In section~3, we review the results of \cite{BFHS04} on the viable parameter region for
this {\it minimal} form of EWBG. In section~4, we turn to the EDMs, and summarize the full set of two-loop contributions to
$d_{\rm Tl}$, $d_n$, $d_{\rm Hg}$, and future observables. Section~5 summarizes our numerical results
on the parameter ranges allowed by the EDMs, and we conclude in Section~6.

\section{Higher-dimensional operators}

We will focus on the SM augmented with the following dimension-six operators in the Higgs sector,
\be 
 {\cal L}_{\rm dim-6} = \frac{1}{\La^2} (H^\dagger H)^3 + \frac{Z^u_{ij}}{\La_{\rm CP}^2} (H^{\dagger} H) U_i^c Q_j H 
 + \frac{Z^d_{ij}}{\La_{\rm CP}^2} (H^{\dagger} H) D_i^c Q_j \tilde{H}  + \frac{Z^e_{ij}}{\La_{\rm CP}^2} (H^{\dagger} H) E_i^c L_j \tilde{H}, \label{ops}
 \ee
 with $\tilde{H} = i \si_2 H^*$.
The first term is required to induce a sufficiently strong first-order transition, while the remaining operators provide the additional source (or sources) of $CP$-violation.  We have introduced two threshold scales for the $CP$-even and $CP$-odd sectors, since they are distinguished according to the
preserved symmetries. However, we will find that they are necessarily of a similar order.

 For the purposes of
 baryogenesis, it would be sufficient to add a single additional complex phase, the most relevant being
 a $CP$-odd coupling of the Higgs to the top,
 \be \label{Zt}
  \frac{{\rm Im}(Z^u_{33})}{\La_{\rm CP}^2} (H^\dagger H) t^c H Q_3.
 \ee
 However, since the threshold scale $\La$ will need to be close to the electroweak scale, it is clearly
 more natural to avoid the introduction of any new source of flavor violation; hence the general
 set of operators in (\ref{ops}), for which we can assume proportionality to the Yukawas,
 \be
  Z^{(u,d,e)}_{ij} = Z^{(u,d,e)} Y^{(u,d,e)}_{ij},
 \ee
 as would be in accordance with the hypothesis of minimal flavor violation (MFV) \cite{mfv}.
 
 There are clearly many other operators allowed by symmetries at this threshold. Focussing for  a moment on the $CP$-odd
 sector, the most dangerous that could arise at a relatively low threshold would be the EDMs and color EDMs of light 
 fermions (the $\theta$-term would be considerably worse, but will be discussed separately below). 
 Generically such operators will only be suppressed by a one-loop factor
 which would be difficult to accomodate for any scenario of EWBG. Our working assumption will be 
 that any new source of $CP$-violation arises only
 from the Higgs sector, e.g. through its coupling to fermions. 
This assumption defers the generation of fundamental EDMs to the two-loop level, 
since the one-loop Higgs-exchange diagrams resulting in EDMs and color EDMs of light fermions 
as well as the tree-level 
 four-fermion contributions to atomic EDMs will be suppressed by the square of the light fermion Yukawa coupling.  
 Thus, the top-operator (\ref{Zt}) maximizes the contribution to EWBG, while affecting EDMs only at two loops. 
Other $CP$-odd dimension-six operators which may be relevant for EWBG, of the form $H^\dagger H W \tilde{W}$ \cite{DHSS91}, 
 are more problematic in this respect, as they mix with the EDM operators at the one-loop level. 
 However, these operators are expected on general grounds to arise only at 
loop level \cite{Arzt:1994gp} from a fundamental theory, and thus are likely to come with correspondingly suppressed coefficients,
making them less attractive for EWBG.

To illustrate this point, we construct a two-Higgs doublet model that provides 
the simplest UV completion for the effective theory considered here. Let us introduce a second (heavy) Higgs doublet $H_h$
with the same hypercharge as $H$. We introduce a non-minimal but flavor-preserving Yukawa sector 
by introducing new complex couplings $c_i$ that are unit matrices in flavor-space,
\be
{\cal L}_{\rm Y} = - Y_u U^c Q (H + c_u H_h)- Y_d D^cQ (\tilde{H} + c_d \tilde{H}_h) - 
Y_e E^c L (\tilde{H} + c_e \tilde{H}_h).
\label{2hdm}
\ee
We also introduce a generic scalar sector,
\be
V_{\rm scalar} = \sum_i \lambda_i (\phi)^4 + \fr 12  m^2 H^\dagger H + \fr 12 m_1^2 H_h^\dagger H_h + (m_{12}^2 H^\dagger H_h+ h.c.),
\label{scalarpot}
\ee
where $\sum_i \lambda_i (\phi)^4$ denotes the set of all gauge-invariant quartic combinations of $H$ and $H_h$.  The effective theory
introduced above is realized  in  the limit where
$m_1^2\gg m^2,~m_{12}^2, ~ v^2$ so that we may integrate out $H_h$. Retaining only the leading order 
terms in $1/m_1^2$ we have,
\be
H_h = -\frac{1}{m_{1}^{2}}(m_{12}^2 H + \lambda H (H^\dagger H)),
\label{H1}
\ee
in which $\lambda$ is the (complex) coefficient in front of $(H^\dagger H)(H_h^\dagger H)$ in (\ref{scalarpot}). 
Substituting $H_h$ from (\ref{H1}) into (\ref{2hdm}) we reproduce the dimension-6 operators in (\ref{ops}) and (\ref{Zt}),
and in particular we find that to leading order in $1/m_1^2$:
\be
\frac{Z_u}{\Lambda_{\rm CP}^2}  \equiv   \frac{\lambda c_u }{m_{1}^{2}};~~~
\frac{Z_d}{\Lambda_{\rm CP}^2}  \equiv   \frac{\lambda^* c_d }{m_{1}^{2}};~~~
\frac{Z_e}{\Lambda_{\rm CP}^2}  \equiv   \frac{\lambda^* c_e }{m_{1}^{2}}.
\ee
The important point in this analysis is that operator (\ref{Zt}) is indeed generated at tree level,
while other potential $CP$-odd operators are loop-suppressed. Also, one may readily notice that both 
$\Lambda$ and $\Lambda_{\rm CP}$ scale in the same way in the limit of large $m_1^2$. One could perform 
the whole analysis of EDMs vs baryon asymmetry in the framework of this particular model (see {\it e.g.} \cite{fhs06}), but we choose
here to study the effective description as it has more generality due to the potentially large 
number of possible UV completions. 

 For the $CP$-even sector, there are also many other operators we should
 include at dimension-six, and as is well known there are quite strong constraints on oblique corrections 
 that would naively require a much larger threshold, of ${\cal O}({\rm few\;TeV})$, than we will consider here. 
 As noted above, the actual tuning of these operators is not prohibitive in this case, relative at least to that required
 for EWBG in the MSSM, and may be ameliorated by further symmetries.  However, whille this is an important issue, 
 our focus will be on exploring whether such a minimal EWBG 
 scenario is viable at all once we impose the full EDM constraints. Our assumption at this point is
essentially that other operators only arise at loop level at this threshold; the question of precisely which
symmetries would ensure this is beyond the scope of the present paper, and we will therefore be prepared to accept a 
certain level of tuning.

\section{Electroweak baryogenesis}
Electroweak baryogenesis relies on a first-order electroweak phase transition as the
source of out-of-equilibrium effects. During the phase transition bubbles of the 
low-temperature (broken) phase nucleate and expand to fill all space.
The $CP$ violating interactions of particles in the plasma with the bubble
wall create an excess of left-handed fermions over the corresponding
anti-fermions. In the symmetric phase the left-handed fermion density biases the
sphaleron transitions to generate a net baryon asymmetry.
To avoid baryon number washout after the phase transition, sphaleron processes
must be sufficiently suppressed within the bubbles. This ``washout criterion" translates
to \cite{M98}
\be \label{wo}
\xi={v_c\over T_c} \ge 1.1,
\ee
and indicates a ``strong'' phase transition. Here $v_c$ denotes the Higgs vev at
the critical temperature $T_c$, where the two minima of the potential become 
degenerate.

\begin{figure}[t] 
\begin{picture}(100,250)
\put(0,20){\centerline{\includegraphics[width=9cm]{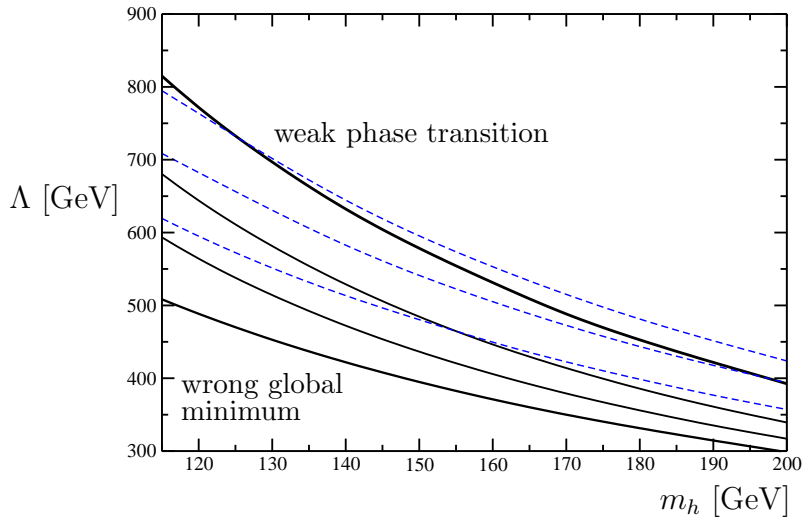}}}
\put(60,120){{$\Lambda$ [GeV]}}
\put(306,6){$m_h~[{\rm GeV}]$}
\put(125,50){\small wrong global}
\put(125,40){\small minimum}
\put(160,145){\small weak phase transition}
\end{picture} 
\caption{\label{param}
The solid lines indicate the strength of the phase transition, with $\xi=1.1,~2,~3$ (from above),
with  $\xi\sim 3$ delineating metastability of the symmetric minimum, while below the fourth (lowest)
solid line the electroweak minimum is no longer the global one. The
dashed lines indicate constant wall thickness $L_w=3T_c^{-1},~6T_c^{-1},~12T_c^{-1}$
(from below).
}
\end{figure}

Including the $(H^\dagger H)^3$ term, the Higgs potential has two free parameters, 
the suppression scale $\Lambda$ of the dimension-six operator and the quartic Higgs 
coupling $\lambda$.  The latter can be eliminated in terms of the physical Higgs mass 
$m_h$, and in this model the Higgs boson is SM-like, so it has to obey the LEP bound 
$m_h>114$ GeV. Note that since the potential is stabilized by the $(H^\dagger H)^3$ term, $\lambda$ can 
be negative, and in this case a barrier in the Higgs potential is present at tree-level, which triggers 
a first order electroweak phase transition \cite{Z93,ZLWY94,GSW04,HO04,BFHS04}.
Here we follow closely the analysis of Ref.~\cite{BFHS04} where, on
computing the 1-loop thermal potential, it was shown 
that the phase transition is strong enough to avoid baryon number washout
if $\Lambda\la820$ GeV. In Fig.~\ref{param} we show the strength of the phase
transition in the $\Lambda-m_h$ plane. The upper solid line delineates the boundary in parameter
space between a strong and weak phase transition. Going to smaller values of $\Lambda$,
the phase transition becomes stronger. As shown in Ref.~\cite{BFHS04}, at around $\xi=3$ the 
symmetric minimum becomes metastable, i.e.~the early Universe would get stuck in
the ``wrong'' vacuum. Finally, below the lowest solid line the electroweak minimum
is no longer the global minimum even at zero temperature. For electroweak baryogenesis,
the interesting region of parameter space lies between the $\xi=1.1$ and $\xi=3$ lines and,  
as in the SM, the phase transition becomes weaker for larger Higgs masses.
Depending on the lower bound on $\Lambda$, Higgs masses up to at least 200 GeV
are compatible with a strong phase transition. For $\Lambda\la400$ GeV, terms suppressed by
higher orders of the cut-off start to become important. In summary, the model allows for a 
strong phase transition in a large part of its parameter space and, as discussed in 
Ref.~\cite{GSW04}, also predicts interesting deviations from the SM Higgs self-couplings, which
which may be measurable at a future linear collider. 

The thickness of the bubble walls, $L_w$, was determined in \cite{BFHS04} and we show dashed lines
of constant $L_wT_c=3,~6$ and 12 in Fig.~1. 
As $\Lambda$ decreases, and the phase transition gets
stronger, the bubbles walls  become thinner. Nonetheless, in a large part of the available
parameter space we have $L_wT_c\gg 1$, i.e. the ``thick wall'' regime, and the wall profile is well-approximated by a hyperbolic tangent, 
$\phi(z)=(v_c/2)(1-\mbox{tanh}(z/L_w))$, where $\phi=\sqrt{2}$Re$(H^0)$.

The dimension-six operators of eq.~(\ref{ops}) induce new sources of $CP$-violation, and 
for baryogenesis the generalized top Yukawa coupling of Eq.~(\ref{Zt}) is the most 
important \cite{ZLWY94}.  We denote the relative phase between this
operator and the ordinary Yukawa interaction, $y_tt^cHQ_3$, by
$\varphi_t={\rm arg}(y_t^*Z_{33}^u)$. Along the bubble wall, the two operators contribute 
with varying weight to the mass of the top. Thus the top mass develops a position dependent
phase $M_t(z)=m_t(z)e^{i\theta_t(z)}$, where
\begin{equation} \label{theta}
\tan\theta_t(z)\approx \sin\varphi_t\frac{\phi^2(z)}{2\Lambda_{\rm CP}^2}\left|\frac{Z_{33}^u}{y_t}\right|.
\end{equation} 
 
Since in this model $L_w\gg T_c^{-1}$, we can treat the interactions between the expanding
bubble wall and the plasma in a WKB approximation, which corresponds to an expansion
in gradients of the bubble profile \cite{JPT95,CJK00}. At first order in gradients a $CP$-violating shift
is induced in the dispersion relations of fermions crossing the bubble wall. For  the top quark one
obtains \cite{kpsw,PSW04,FH06},
\begin{equation} \label{disp}
E=E_0\pm\Delta E=\sqrt{p^2+m_t^2}\mp {\rm sign}(p_z)\theta_t'
\frac{m_t^2}{2\sqrt{p^2+m_t^2}\sqrt{p_z^2+m_t^2}},
\end{equation}
where the upper (lower) sign is for particles (anti-particles). Thus particles and antiparticles
experience a different force when they cross the bubble wall. These forces 
generate $CP$-violating source terms in the Boltzmann equations that describe transport
processes in the hot plasma. The source term generated by the top quark dominates
baryon number production, while source terms of the light fermions are suppressed by
$(m/m_t)^4$. 

We compute the baryon asymmetry using the formalism of Ref.~\cite{FH06}, where the full transport
equations and the values of interaction rates can be found. The transport equations are dominated
by the left- and right-handed top quarks, while the Higgs bosons are a subleading 10\% effect. 
As in  Ref.~\cite{FH06}, we keep the $W$-scatterings at a finite rate and include the position 
dependence of the thermal averages. 
For each parameter combination, $\Lambda$ and $m_h$, we compute the bubble wall 
properties $\xi$ and $L_w$. Together with $\Lambda_{\rm CP}$ and $\varphi_t$ they
enter the dispersion relation and determine the baryon asymmetry, which also
depends weakly on the bubble wall velocity $v_w$. Varying
$v_w$ between $10^{-3}$ and 0.5 changes the baryon asymmetry by only 20\% \cite{FH06}, and
we use $v_w=0.1$ in our evaluations.

This model can actually generate the observed baryon asymmetry for a relatively wide
range of paramaters \cite{FH06}, with the asymmetry naturally increasing for decreasing 
$\Lambda$. The $CP$-violating part of the dispersion relations scales as  $\xi^4$ and thus is enhanced 
by the strength of the phase transition. The resulting baryon asymmetry is
proportional to the $CP$-violating parameter $\varphi_t/\Lambda_{\rm CP}^2$, and we will
now confront the corresponding value required for successful baryogenesis with
the constraints imposed by the EDMs.

\section{Electric dipole moments}

In this section we will first summarize the relevant formulae for the observable EDMs in terms of
the $CP$-odd operators normalized at 1 GeV, including a brief discussion of the observables to be probed in the next generation of
experiments.  We then turn to a discussion of the various two-loop
contributions which arise at leading order in the present framework.

\subsection{Observable EDMs}

We first recall the most
significant flavor-diagonal $CP$-odd operators at 1 GeV (see \cite{PRrev} for a recent review). 
Up to dimension six, the
corresponding effective Lagrangian takes the form,
\begin{eqnarray}
\label{leff}
{\cal{L}}_{eff} &=& \frac{g_s^2}{32\pi^{2}}\ \bar\theta\
G^{a}_{\mu\nu} \widetilde{G}^{\mu\nu , a} \nonumber\\  && -
\frac{i}{2} \sum_{i=e,u,d,s} d_i\ \overline{\psi}_i (F\sigma)\gamma_5 \psi_i  -
\frac{i}{2} \sum_{i=u,d,s} 
\widetilde{d}_i\ \overline{\psi}_i g_s (G\sigma)\gamma_5\psi_i \nonumber\\
  && + \frac{1}{3} w\  f^{a b c} G^{a}_{\mu\nu} \widetilde{G}^{\nu \beta , b}
G^{~~ \mu , c}_{\beta}+ \sum_{i,j=e,q} C_{ij} (\bar{\ps}_i \ps_i) (\ps_j i\gamma_5 \ps_j)+ \cdots
\end{eqnarray}
Since we require a large phase in the top sector, the only reasonable strategy to avoid the strong $CP$ problem is to
invoke the axion mechanism \cite{PQ}, which removes the $\theta$-term from the above list of operators, and we will
adopt this approach here. In addition, as discussed below, the four-fermion operators are generically subleading, and will be ignored 
for most of our numerical analysis. They are included in our discussion below for completeness, because they
actually arise at tree-level, and can provide a significant contribution if there is a mild hierarchy in the coefficients of
the operators in the up and down sectors, as would arise for large $\tan\beta$ in a 2HDM for example.

The physical observables can be conveniently
separated into three main categories, depending on the physical mechanisms via which an
EDM can be generated: EDMs of paramagnetic 
atoms and molecules, EDMs of diamagnetic atoms, and the neutron EDM. The current constraints
within these classes are listed in Table~1.

\begin{table}
\begin{center}
\begin{tabular}{||c|c|c||}
 \hline
  Class & EDM & Current Bound  \\
  \hline
  Paramagnetic & $^{205}{\rm Tl}$ & $|d_{\rm Tl}| < 9 \times 10^{-25} e\, {\rm cm}$ \cite{Tl}  \\
  Diamagnetic & $^{199}{\rm Hg}$ & $|d_{\rm Hg}| < 2   \times 10^{-28}  e\, {\rm cm}$ \cite{Hg}  \\
  Nucleon & $n$ & $|d_n|  <  3\times 10^{-26} e\, {\rm cm}$ \cite{n} \\
  \hline
\end{tabular}
\caption{Current constraints within the three representatve classes of EDMs}
\end{center}
\label{explimit}
\end{table}

\subsubsection{EDMs of paramagnetic atoms -- thallium EDM}

Among various paramagnetic systems, the EDM of 
the thallium atom currently provides the best constraints. Atomic calculations
summarized in \cite{fgreview} link the atomic EDM with $d_e$ and various $CP$-odd electron-nucleon interactions, of which
we shall only consider the most relevant, namely $C_S\bar e i \gamma_5 e \bar NN$,
\be 
d_{\rm Tl} = -585 d_e -   e\ 43 ~{\rm GeV} C_S^{\rm singlet}.
\label{dtl}
\ee
The relevant atomic matrix elements are known to a precision of 
$10-20\%$. For completeness, although not required for the following analysis, 
we present the dependence of $C_S$ on the four fermion sources $C_{ie}$, for $i=d,s,b$ \cite{dlopr},
\be
C_S^{\rm singlet} = C_{de}\frac{29~{\rm MeV}}{m_d}+ C_{se}\frac{\kappa \times 220~{\rm MeV}  }{m_s}+ 
C_{be}\frac{66~{\rm MeV}(1-0.25 \kappa )}{m_b},
\ee
where $\kappa \equiv \langle N|m_{s}\overline{s}s|N\rangle /220$ MeV$ \sim 0.5 - 1.5$.

\subsubsection{Neutron EDM}

The neutron EDM  $d_n$ plays a crucial role in constraining $CP$-odd sources in the quark sector,
and the corresponding bound has recently been lowered by a factor of two. We will make use of the results obtained 
using QCD sum rule techniques \cite{PR,DPR} (see
\cite{CDVW} for alternative chiral approaches), wherein under Peccei-Quinn relaxation the
contribution of sea-quarks is also suppressed at leading order \cite{PR,DPR}:
\ba
 d_{n}(d_q, \tilde d_q) &=& (1.4 \pm 0.6)(d_d-0.25d_u) + (1.1 \pm 0.5)e(\tilde d_d + 0.5\tilde d_u) \nonumber\\
  &&+ 20\,{\rm MeV}\times e~ w +{\mathcal O}(C_{qq}).
\label{dn1}
\ea
The quark vacuum condensate, $\langle \bar qq\rangle = (225 \, {\rm MeV})^3$, 
has been used in this relation -- the proportionality to $d_q\langle \bar qq\rangle \sim 
 m_q\langle \bar qq\rangle \sim f_\pi^2m_\pi^2$ removes 
any sensitivity to the poorly known absolute value of the light quark masses. 
Here $\tilde d_q$ and $d_q$ are to be normalized at the hadronic scale 
which we assume  to be 1 GeV. 

The contribution of the Weinberg operator is known to less precision than the quark (C)EDMs (a factor of 2--3), but 
is included here only for completeness, as it provides a negligible contribution in the present scenarios. There are additional 
four-quark contributions that are also unimportant here.

\subsubsection{EDMs of diamagnetic atoms -- mercury EDM}

Constraints on the EDMs of diamagnetic atoms are also powerful probes; the current limit on the EDM of mercury 
\cite{Hg} stands as one of the most sensitive constraints on new $CP$-odd phases.
The atomic EDM of mercury arises from several important sources (see e.g. \cite{KL}), namely, 
the Schiff moment $S$ \cite{Schiff} of the nucleus, the electron EDM $d_e$, and also 
various electron-nucleon and nucleon-nucleon interactions.
The important contributions here arise from the Schiff moment, which depends primarily on the 
quark CEDMs via $CP$-odd pion nucleon couplings: $S=S[\bar{g}_{\pi NN}(\tilde{d}_u,\tilde{d}_d)]$, and also
the electron EDM.

Combining the atomic $d_{\rm Hg}( S)$ \cite{Hgnew}, nuclear $S(\bar g_{\pi NN})$ \cite{DS}, and QCD $\bar{g}^{(1)}_{\pi NN}(\tilde{d}_q)$ \cite{P},
components of the calculation, we have
\be
d_{\rm Hg} = 7\times 10^{-3}\,e\,(\tilde d_u - \tilde d_d)  + 10^{-2}\, d_e + {\mathcal O}(C_S,C_{qq})
\label{Hgmaster}
\ee
where the overall uncertainty is rather large, a factor of 2-3, due to significant cancelations between various contributions.
As noted above, additional contributions from various four-fermion operators  have been suppressed.
In practice, the most valuable feature of $d_{\rm Hg}$ is its sensitivity to the triplet combination of
CEDM operators $\tilde d_i$.

\subsubsection{Future experimental sensitivity}

The experimental situation is currently very active, and a number of new EDM experiments 
promise to improve the level of sensitivity by one to two orders of magnitude in each characteristic class in the coming 
years (see e.g. \cite{oprs}). Beyond the ongoing experiments, these comprise searches for EDMs of polarizable paramagnetic 
molecules \cite{PbO,YbF} and solid state systems \cite{Lamoreaux}, which are primarily sensitive to the electron EDM and aiming 
at a sensitivity of $10^{-29}\,e$cm, new searches for the neutron EDM in cryogenic systems \cite{lansce} with sensitivity goals
of $10^{-28}\,e$cm, and searches for nuclear EDMs using charged nuclei in storage rings \cite{bnl,lopr}.

\subsection{Contributions from the Higgs sector}

\subsubsection{Two-loop Contributions}

The loop contributions to EDMs in this scenario are very similar to those present in a 2HDM, from which it can clearly
be obtained by integrating out the heavy Higgses. Thus the loop functions that appear are those computed by Barr and Zee
in the latter case \cite{BZ} (see also \cite{cky}). 

\begin{figure}
\centerline{\includegraphics[bb=130 370 470 720, clip=true,width=5cm]{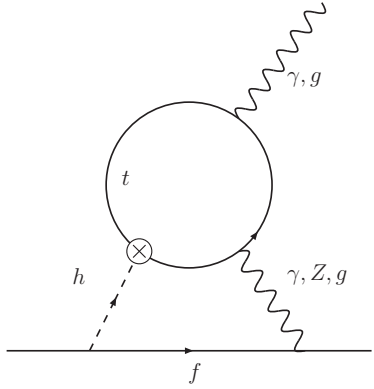}}
\caption{\footnotesize The two-loop contribution to $d_f$ and $\tilde{d}_f$ mediated by an induced pseudoscalar
$hF\tilde{F}$ coupling generated by the top-loop. \label{diagram1}}
\end{figure}

\begin{figure}
\centerline{\includegraphics[bb=0 400 700 710, clip=true,width=10cm]{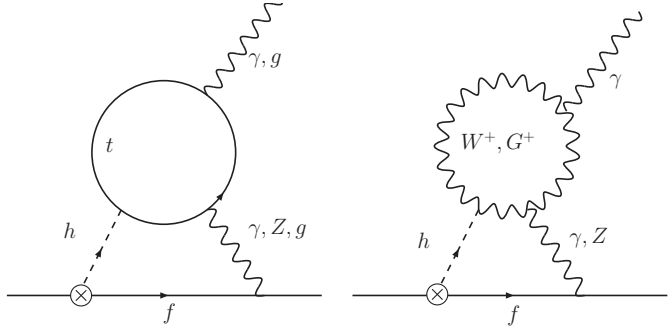}}
\caption{\footnotesize The two-loop contributions to $d_f$ and $\tilde{d}_f$ mediated by an induced scalar
$hFF$ coupling, generated either by quark (top) loops  or various vector boson and/or Goldstone loops. \label{diagram2}}
\end{figure}

To summarize the results, its convenient to first focus on $d_f$, and to split the contributions into those 
that arise via an effective pseudoscalar $hF\tilde{F}$ vertex and those arising from the scalar $hFF$ vertex.
The generalization to consider $hZ\tilde{F}$ and $hZF$ vertices is then straightforward, although in fact 
the $Z$-mediated contributions are highly suppressed for $d_e$. The diagrams in each class are shown schematically in 
Fig.'s \ref{diagram1} and \ref{diagram2}. A few remarks on the relevant diagrams are in order:
\begin{itemize}
\item For the $hF\tilde{F}$-mediated contributions, Yukawa suppression allows us to limit attention to the
top-loop. Note also that the only pseudoscalar effective vertices are $h\ga\ga$ and $h\ga Z$. 
$W^+$ and its Goldstone component $G^+$ do not contribute here as $CP$-violation only enters the
neutral Higgs sector. 
\item For the $hFF$-mediated contributions, $CP$-violation enters on the external fermion line, and so 
more modes may propagate in the internal loop; we should allow in general for $t$, $W$, and $G^+$, the latter
two in various combinations.
\end{itemize}

Consequently, the relevant fermion EDMs can be decomposed as follows,
\be
 d_f = d_f^{\ga\ga t} + d_f^{\ga Z t} + d_f^{\ga\ga W}, 
\ee
where the first two terms refer to the top-loops in Figs.~2 and 3, while the third refers to the boson loop. The color EDMs of quarks are predominantly
generated via the top-loop, 
\be
 \tilde{d}_q = d_q^{ggt}.
\ee 

The explicit expressions for the  individual contributions are very similar to those arising in the 2HDM, but there are some differences,
and we will present the full results for completeness. The loop functions needed are those of Barr and Zee, 
\ba
 f(z) &=& \frac{z}{2} \int_0^1 dx \frac{1-2x(1-x)}{x(1-x)-z} \ln \left(\frac{x(1-x)}{z}\right), \\
 g(z) &=& \frac{z}{2} \int_0^1 dx \frac{1}{x(1-x)-z} \ln \left(\frac{x(1-x)}{z}\right), 
\ea
which satisfy $f(1)=1/2$, $g(1)=1$, and have the asymptotics, $f\sim (1/3)\ln z$ and $g\sim (1/2)\ln z$ at large $z$.
The loop function $f$ arises from the effective $hFF$ vertex, and $g$ from the $hF\tilde{F}$ vertex.

In terms of these functions, we have
\be
 \frac{d^{\ga\ga t}_f}{e} = -Q_f \frac{\al}{6\pi^3} \frac{m_f}{\La^2} 
{\rm Im}\left[Z^u g(m_t^2/m_h^2) - Z^f f(m_t^2/m_h^2)\right],
\ee
where $Q_f$ is the electric charge of the fermion $f$. 
The analogous results generated by the $hZF$ and $hZ\tilde{F}$ vertices follow on inserting
the corresponding vector components of the $Zf$ couplings:
\be
 \frac{d^{\ga Z t}_f}{e} = -\frac{\left(\mp1/4+Q_f\sin^2\theta_W\right)\left(-1/4+2\sin^2\theta_W/3\right)}{\sin^2\theta_W\cos^2\theta_W}\frac{\al}{4\pi^3} 
 \frac{m_f}{\La^2} {\rm Im}\left[Z^u \tilde{g}(z_h,z_Z) - Z^f \tilde{f}(z_h,z_Z)\right],
\ee
where the $(\mp)$ refers to (up/down)-type vertices, $z_h=m_t^2/m_h^2$, $z_Z=m_t^2/m_Z^2$, and the 2-parameter loop functions 
are given by,
\be
 \tilde{X}(x,y) = \frac{yX(x)}{y-x} + \frac{xX(y)}{x-y}, \;\;\;\; {\rm with} \; X=f,g.
\ee
This correction for $d_e$ is negligible, as it is within the 2HDM \cite{BZ}, but the correction for the quark EDMs is on the order 
of 30-40\%.

The expressions for the color EDMs of quarks follow similarly,
\be
 \frac{d^{gg t}_q}{g_s} = \frac{\al}{16\pi^3} \frac{m_f}{\La^2} 
 {\rm Im}\left[Z^u g(m_t^2/m_h^2) - Z^ff(m_t^2/m_h^2)\right],
\ee
and will generically provide the largest contribution to the hadronic EDMs, which will in turn provide the most stringent constraints
on the scenarios considered here.

For the contributions mediated by the scalar effective vertices, $hFF$ and $hZF$ etc., since $CP$-violation
enters on the external fermion line, we should also consider possible boson internal loops. We will take only the largest of these into
account associated with a $W$-loop,
\be
 \frac{d^{\ga\ga W}_f}{e} = -Q_f \frac{ 3\al}{16\pi^3} \frac{m_f}{\La^2} 
{\rm Im}Z^f f(m_W^2/m_h^2),
\ee
while additional contributions from the Goldstone components, and internal $Z$ lines are considerably smaller.

\subsubsection{Subleading Contributions}

The Weinberg operator is also generated by similar two-loop diagrams. However, it is relatively small in this scenario where
the $CP$-odd phase is limited to the neutral Higgs sector. It is generated predominantly in the 2HDM through charged Higgs contributions.

The complex corrections to the Yukawa couplings can of course also generate $CP$-odd four-fermion operators via tree-level
Higgs exchange. We have ignored these because, despite being generated at tree-level, their contribution to the observable EDMs
is still generically suppressed relative to the fermion EDMs by an order of magnitude.
However, this conclusion may not hold if the model has a non-generic normalization for the operators in the two isospin sectors, e.g. if it 
were to arise from a 2HDM, there is the possibility for $\tan\beta$-enhancement of these 4-fermion contributions \cite{barr}.

\section{Numerical constraints}

We will consider a couple of scenarios in presenting the EDM constraints on the parameter space. 

\subsection{Single threshold}

If we assume that the new $CP$-even and $CP$-odd physics lies at around the same threshold scale, we can set $\La=\La_{\rm CP}$,
and exhibit contours on the remaining two-dimensional $(\La,m_h)$ parameter space. This is presented in Fig.~4, where three
$\eta_b$ contours are contrasted with bounds from the Tl, Hg, and neutron EDMs. The contours of $\eta_b$ are labelled in units of
the experimental value, taken to be \cite{wmap}
\be
 \et_b =\frac{n_b}{s} = 8.9 \times 10^{-11}.
\ee
The EDM contours are set to {\it twice} the existing 1$\si$ experimental bound.
This reflects the existing estimates for the theoretical
precision in these calculations, and we will interpret these contours as 1$\si$ exclusions in parameter space. We make use of the 
standard anomalous dimensions to run these operators down to 1 GeV, having set Im$(Z)=\pm 1$ at the threshold in all  cases.
This assumes, as discussed in Section~2, that these operators are generated at tree-level within an appropriate UV completion.
We can limit our attention to this scenario, as a loop suppression factor, $Z \sim 1/(16\pi^2)$, while clearly ameliorating the EDM
constraints on the threshold, would not allow for the generation of the required baryon asymmetry unless the thresholds are too low,
i.e. at or below $v_{\rm EW}$.

\begin{figure}
\begin{picture}(450,220)
\put(0,0){\centerline{\includegraphics[width=8.3cm]{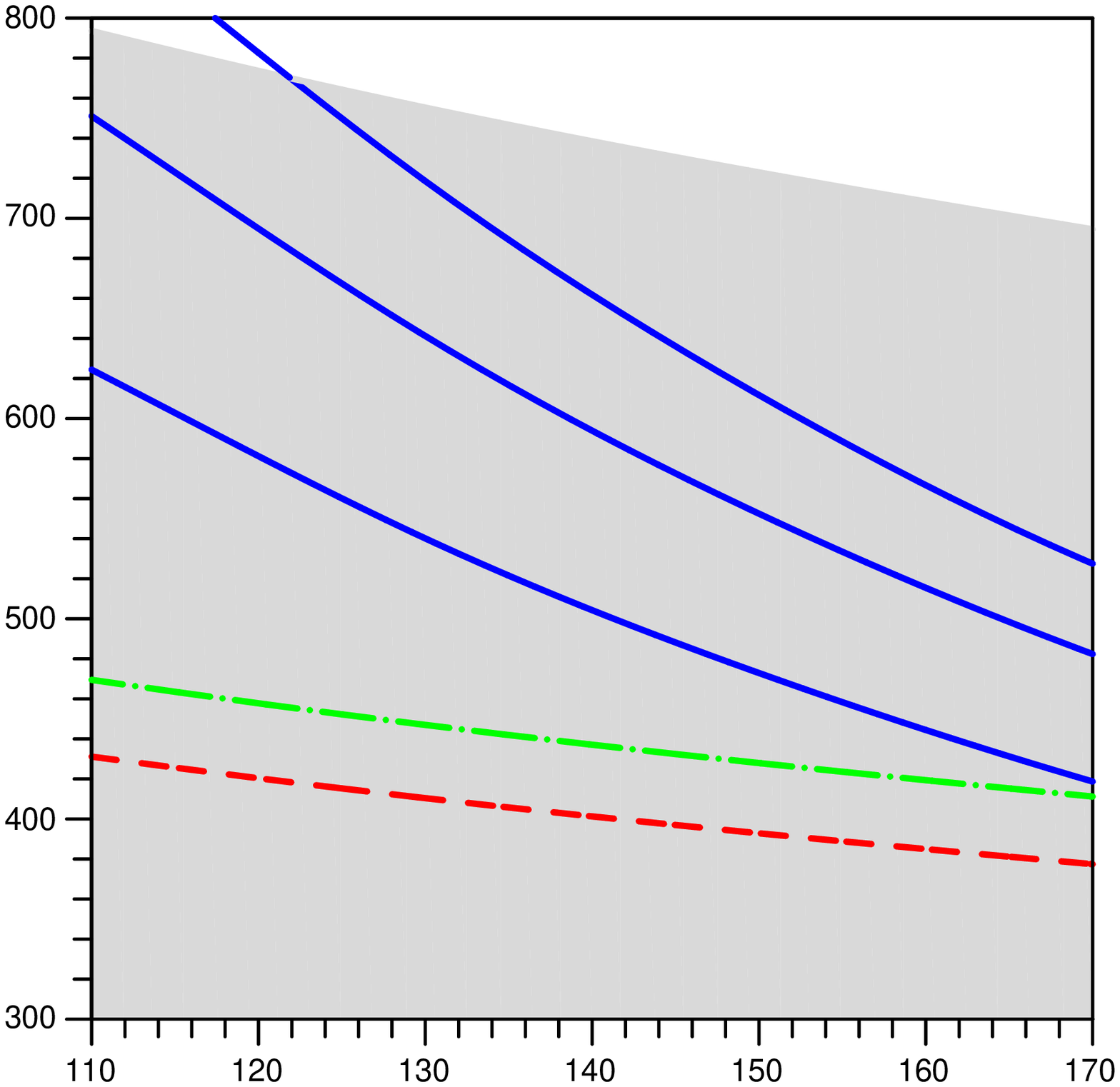}\hspace*{0.0cm}\includegraphics[width=8.3cm]{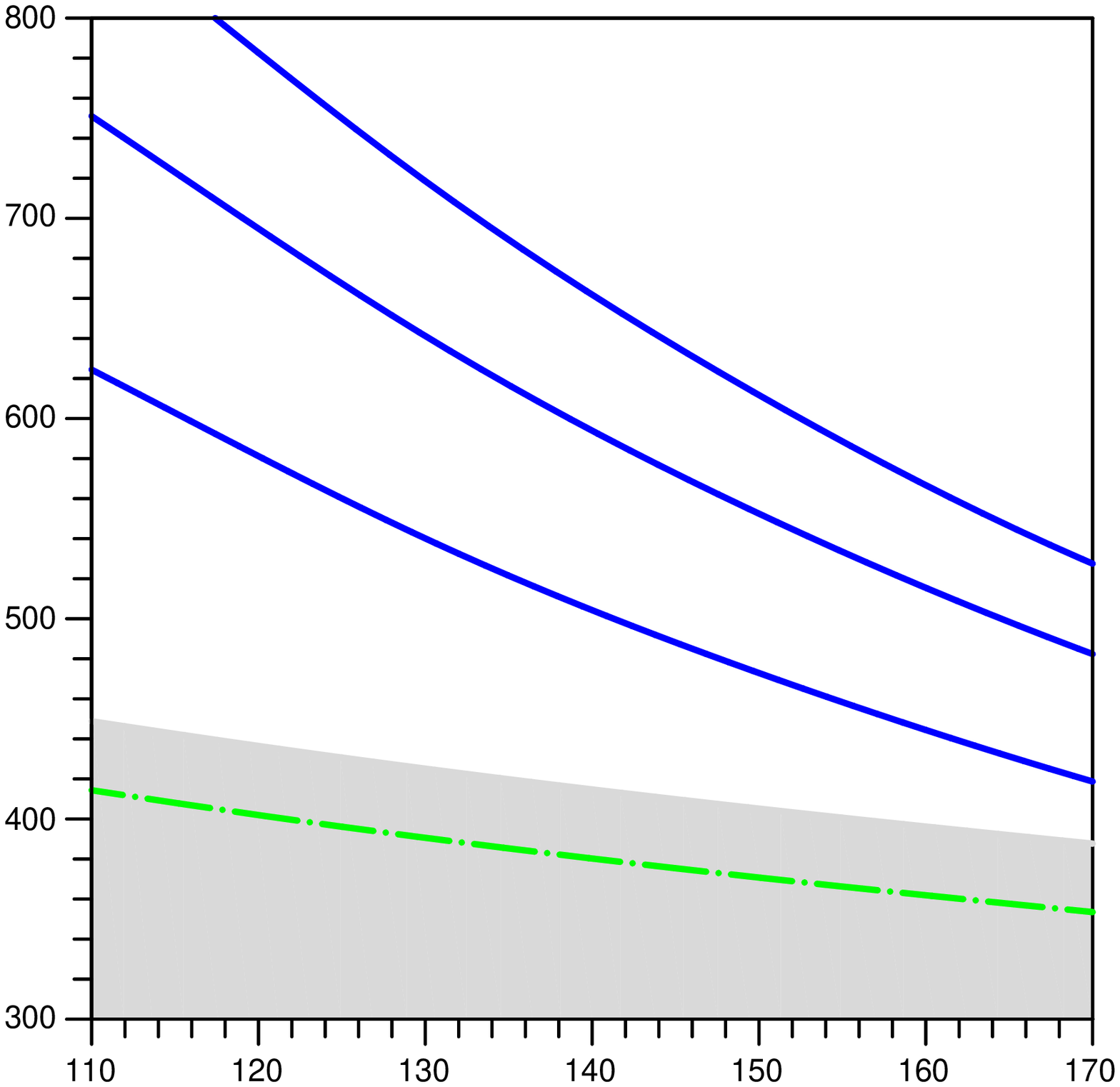}}}
\put(163,0){{$m_h$ [GeV]}}
\put(15,220){{$\Lambda$ [GeV]}}
\put(25,60){\small {$d_{Hg}$}}
\put(25,95){\small {$d_{Tl}$}}
\put(25,200){\small {$d_{n}$}}
\put(185,80){\small {$\et_{10}$}}
\put(185,105){\small{$\et_1$}}
\put(185,125){\small{$\et_{0.1}$}}
\put(400,0){{$m_h$ [GeV]}}
\put(250,220){{$\Lambda$ [GeV]}}
\put(260,50){\small {$d_{Tl}$}}
\put(260,85){\small {$d_{n}$}}
\put(420,80){\small {$\eta_{10}$}}
\put(420,105){\small{$\eta_1$}}
\put(420,125){\small{$\eta_{0.1}$}}
\end{picture}
\caption{\footnotesize Contours of $\eta_b$ -- labelled as $\et_x$ where $\eta_b/\eta_{\rm exp}=x$ -- and the EDMs over the $\La$ vs $m_h$ plane, with 
correlated thresholds, $\La_{\rm CP}=\La$. The shaded region is excluded by the EDMs, primarily the neutron EDM bound in this case.
On the left, we retain only a single $CP$-odd phase in the top-Higgs vertex, while on the right the full set required to retain the Standard Model
flavor structure is introduced, which allows the $d_n$ and $d_{Hg}$ bounds to be weakened (the $d_{Hg}$ contour actually lies below the 300 GeV ctuoff
imposed on $\La$ and so does not appear on the right-hand plot).}
\end{figure}

On the left of Fig.~4, we consider the minimal scenario required for EWBG, namely a single additional $CP$-odd phase in the top-Higgs
vertex. We see that the (recently updated) neutron EDM bound provides the strongest constraint, with the shaded region below the contour covering
all of the viable parameter space. However, if we enforce the constraint of having no new flavor structure, additional $CP$-odd sources are
allowed, which necessarily allows for partial cancelations. Indeed, its clear in this case that there are a sufficient number of parameters available  to 
cancel the contributions of Im$(Z^u_{33})$ -- the phase relevant for baryogenesis -- to the observable EDMs. This would be a tuned situation of course, and rather than 
map out the possible cancelations, we choose to consider just a generic example, given that the motivation to include additional $CP$-odd sources
was to retain the structure of minimal flavor violation, and not to specifically consider the possibility of fine-tuning away the EDM constraints. With this 
viewpoint in mind, on the right of Fig.~4, we consider a characteristic example with Im$(Z^u)$=Im$(Z^d)$=Im$(Z^e)$=1. Partial cancelations,
on the order of 50-70\% in this mass range, are then sufficient to open up a significant allowed region where EWBG is viable, which interestingly tends to favor the region
of low Higgs mass. The cancelations in this case predominantly affect the
hadronic EDMs, $d_n$ and $d_{Hg}$ which are primarily determined by the quark color EDMs, while $d_{Tl}$ is less affected due to the additional
impact of the $W$-loop . This illustrates the generic complementarity of the constraints, so that while partial cancelations may be quite generic, there
is a limit to the suppression of the EDMs that may be achieved in this way, without very precise fine-tuning of a number of parameters.

Its worth noting here that the sign of the induced EDMs is actually predicted if there is a single $CP$-odd source, namely the top-Higgs coupling $Z^u_{33}$.
Reproducing the correct baryon asymmetry requires that Arg($y_t^* Z^u_{33})<0$, so the induced neutron EDM, for example, would be negative. Of course,
in practice, it would be difficult to separate such a minimal scenario from the more generic case with multiple $CP$-odd sources without further input
from multiple EDM measurements.

\subsection{Decoupled thresholds}

On general grounds, it is more natural to decouple the two thresholds. To present the results, its convenient to fix the $CP$-odd threshold
$\La_{\rm CP}$ by imposing the required value for $\et_b$ given a choice of $\La$ and $m_h$. We again minimize the constraints by taking  Im$(Z^u)$=Im$(Z^d)$
which allows for partial cancelations,
 and in Fig.~5 exhibit the 
resulting plots of $d_n$ versus $m_h$
for various values of $\La$. Similar, but slightly less constraining, plots can be obtained for the other EDMs. Note that the falloff of $\et_b$ with
$m_h$ is primarily the reason for the steep rise of the EDM  for larger Higgs masses, due to the need to lower  
$\La_{\rm CP}$ as $\eta_b$ decreases. The curves shown 
indeed do not extend much further to the right before $\La_{\rm CP}$ becomes too low for the EFT treatment to be reliable. 
There is also an excluded region in the lower section of the plot due to the fact that the phase transition becomes stronger 
for lower $m_h$, and ultimately the symmetric vacuum becomes metastable (see Fig.~1). Our thick-wall approximation actually breaks down somewhat before
this point, and so the lower excluded region extends slightly further in Higgs mass than is apparent from the metastability line in Fig.~1.  This constraint, in concert 
with the EDM bound, ensures that the plot exhibits
a quite precisely defined viable region, bounded for low Higgs mass by the direct search bound, and for large Higgs mass by the EDMs, while also requiring
a minimal value for the EDMs (in the absence of fine-tuning) which is less than an order of magnitude below the current sensitivity.\footnote{Similar results were found for electroweak
baryogenesis in the 2HDM \cite{fhs06}.}  The
allowed range for the $CP$-even threshold is limited to:
\be
  400\;{\rm GeV} < \La < 800\;{\rm GeV}.
\ee
As compared to the plots with correlated thresholds, this result does not dramatically
alter the conclusions. This can be understood by taking a closer look at the $CP$-odd threshold scale within the viable region of Fig.~5,
and noting that it does not differ significantly from the value of $\La$. For example,
for $\La=500$~GeV, within the viable range $\La_{\rm CP}$ varies from about 900~GeV at $m_h\sim 150$~GeV to around 400~GeV at
$m_h\sim 165$~GeV. The variation for other values of $\La$ is similar.

\begin{figure}
\begin{picture}(450,220)
\put(0,0){\centerline{\includegraphics[width=8.3cm]{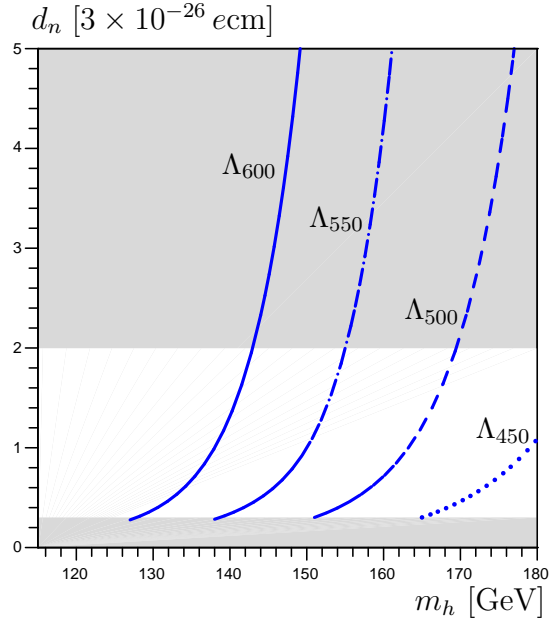}}}
\put(280,0){{$m_h$ [GeV]}}
\put(135,220){{$d_n\; [3\times 10^{-26}\, e{\rm cm}]$}}
\put(207,165){\small {$\La_{600}$}}
\put(240,145){\small {$\La_{550}$}}
\put(276,110){\small {$\La_{500}$}}
\put(303,65){\small {$\La_{450}$}}
\end{picture}
\caption{\footnotesize Fixing several values of $\La$ --  labelled as $\La_x$ where $\La=x$ [GeV] --  $d_n$ is plotted against $m_h$, with $\La_{\rm CP}$ fixed to ensure that $\eta_b$ matches its observed value. Note that the EDMs are only logarithmically dependent on $m_h$, and thus the primary dependence arises implicitly via $\eta_b$. The shaded 
upper region is excluded by the neutron EDM bound, while the shaded lower region is excluded by metastability of the symmetric vacuum.}
\end{figure}

\section{Concluding Remarks}

Electroweak baryogenesis remains an interesting mechanism for many reasons, combining as it does various subtle features of the electroweak sector
of the Standard Model with rather minimal new physics input. The fact that it is under strain from our current knowledge of the electroweak sector and
existing EDM bounds, only serves to emphasize that since it relies on weak scale physics it is genuinely testable, in contrast to high-scale mechanisms
such as leptogenesis. 

In this paper, we discussed the current status of the EDM constraints on perhaps the minimal EWBG scenario where the required new physics emerges
purely from higher-dimensional operators in the Higgs sector. The situation is interesting as the existing constraints, while strong, still allow a reasonable
range for the new thresholds, particularly with a light Higgs. Furthermore, 
the predictions for the level of sensitivity attainable in the next-generation of EDM experiments has profound implications for these scenarios.
If, for the moment, we lock $\La_{\rm CP}=\La$, then the sensitivity attainable in searches for the electron and neutron EDMs would correspond
to a threshold sensitivity of
\be
 \La_{\rm CP} \sim 3\; {\rm TeV},
\ee
over the relevant Higgs mass range, which is well beyond the viable region of (untuned) parameter space for this mechanism of 
EWBG. The sensitivity of the proposed search for the deuteron EDM is even more impressive, with sensitivity up to 30 TeV.
Thus, even with a conservative treatment of the EDM precision, it seems clear that EWBG as realized in the form considered here will be put to the 
ultimate test with the next generation of experiments.

\subsection*{Acknowledgements}
The work of M.P. and A.R. was supported in part by NSERC, Canada. Research at the Perimeter Institute 
is supported in part by the Government
of Canada through NSERC and by the Province of Ontario through MEDT.

\end{document}